\pdfoutput=1
\documentclass[11pt]{article}

\usepackage[preprint]{acl}

\usepackage{times}
\usepackage{latexsym}
\usepackage{times}
\usepackage{amsfonts,amssymb}
\usepackage{bbding}
\usepackage{caption}
\usepackage{subcaption}
\usepackage{xcolor}
\usepackage{tikz}

\usepackage[T1]{fontenc}
\usepackage[utf8]{inputenc}

\usepackage{microtype}

\usepackage{inconsolata}

\usepackage{graphicx}

\title{User Feedback Alignment for LLM-powered Exploration \\in Large-scale Recommendation Systems }

\author{
  Jianling Wang$^{1*}$, Yifan Liu$^{2*}$, Yinghao Sun$^3$, Xuejian Ma$^2$, Yueqi Wang$^2$, \\% Line break after 6 authors
  \bf{He Ma$^2$, Zhengyang Su$^2$, Minmin Chen$^1$, Mingyan Gao$^2$,}\\ \bf{Onkar Dalal$^2$, Ed H. Chi$^1$, Lichan Hong$^1$,  Ningren Han$^2$, Haokai Lu$^1$} \\ % Second line of authors
  $^1$Google DeepMind \quad $^2$YouTube \quad $^3$Google Labs\\ % Affiliations on one line, separated by space (\quad or \qquad)
  \texttt{\{jianlingw, yifanliu, sunmo, xuejianma, yueqiw, htm, susteven,} \\ % Break email list for readability
  \texttt{minminc, mingyan, onkardalal, edchi, lichan, peterhan, haokai\}@google.com} % Complete the email list
}

\begin{document}
\maketitle
\def\thefootnote{*}\footnotetext{indicates equal contribution}
\begin{abstract}

%User interest exploration in large-scale recommendation systems is challenging due to the influence of the system feedback loop and the lack of signals on users' exploration patterns. 
% Exploration aiming to broaden user experiences and potentially uncover hidden preferences is hard to solve problem in large-scale recommendation systems due to the influence of the system feedback loop and the lack of signals on users' exploration patterns. 
% Large language models (LLMs) have shown potential in solving this problem by leveraging their world knowledge and reasoning abilities to recommend novel contents that are outside of the system feedback loop. A key challenge in improving LLM-based exploration is how to align LLMs with user preferences while preserving LLMs' own knowledge and reasoning capabilities.
% This paper introduces a novel approach that combines the hierachical planning paradigm with LLM inference-time scaling to improve the recommendation relevancy without compromising novelty. While using LLM to plan for the next novel user interest, we decouple novelty and user-alignment objectives and train two LLMs focusing on each objective respectively. We then scale up the novelty-focused LLM's inference and select the best-of-n predictions using the user-aligned LLM. Live experiments on a large-scale recommendation system demonstrate the efficacy of our method, showing significant gains in both user satisfaction (measured by watch activity and active user counts) and exploration diversity.

Exploration, the act of broadening user experiences beyond their established preferences, is challenging in large-scale recommendation systems due to feedback loops and limited signals on user exploration patterns. Large Language Models (LLMs) offer potential solutions by leveraging their world knowledge to recommend novel content outside these loops. A key challenge is aligning LLMs with user preferences while preserving their knowledge and reasoning. To enhance planning for new user interests using LLMs, this paper introduces a novel approach that combines hierarchical planning with LLM inference-time scaling. It aims to improve recommendation relevancy without compromising novelty. We decouple novelty and user-alignment, training separate LLMs for each objective. We then scale up the novelty-focused LLM's inference and select the best-of-n predictions using the user-aligned LLM. Live experiments demonstrate efficacy, showing significant gains in both user satisfaction (measured by watch activity and active user counts) and exploration diversity.
\end{abstract}

\section{Introduction}

Large Language Models (LLMs) present a significant opportunity to revolutionize recommendation systems \cite{wu2024survey}, due to their powerful reasoning, planning, and world knowledge capabilities. Traditional recommendation backbones, such as collaborative filtering and content-based methods, typically suggest items by identifying similar users based on past interactions, which often reinforce existing preferences and perpetuate feedback loops \cite{chaney2018algorithmic, mansoury2020feedback}. LLMs can overcome these limitations by leveraging their vast world knowledge to generate novel and diverse recommendations that go beyond a user's historical interactions, thus driving long-term user engagement \cite{chen2021exploration}.

% Building on the exciting potential of LLMs for recommendation systems \cite{bao2023tallrec,lin2023rella,wang2024large,christakopoulou2023large}, recent advancements offer promising avenues to overcome feedback loops and enhance recommendation exploration.
Among recent advancements leveraging LLMs for recommendation systems \cite{bao2023tallrec,lin2023rella,wang2024large},
the hierarchical planning paradigm \cite{wang2024llms} stands out as a promising and \textit{deployable} approach that combines an LLM, which provides high-level guidance, with traditional recommenders for efficient item-level serving. As this solution has been adopted in industry, the subsequent challenge lies in effectively integrating real-world human feedback into the LLM. While human feedback is key to optimizing LLMs \cite{ouyang2022training}, systematically incorporating it into recommendation systems remains an under-explored area, offering both challenges and opportunities for future research.

% \jw{A primary obstacle is effectively utilizing real-world human feedback to refine LLM-driven recommendations. Methodologies such as Reinforcement Learning from Human Feedback (RLHF) \cite{ouyang2022training} have demonstrated success in other domains, relying on explicit comparative judgments (e.g., side-by-side comparisons). However, recommendation systems present a distinct context. Explicit expressions of user preference, such as ratings or reviews, are often scarce. Instead, the system relies predominantly on implicit signals, like clicks, dwell time, and other behavioral data, which are noisy and only indirectly reflect the actual user satisfaction. The challenge lies in developing methods to effectively extract meaningful preference signals from this noisy implicit feedback and translate them into a robust training objective for LLMs, enabling them to learn and align with users' underlying desires more accurately.}

Using real-world human feedback is challenging because recommendation systems rely on noisy implicit signals (e.g., clicks or dwell time) instead of explicit comparative judgments (e.g., side-by-side comparisons). This makes it hard to translate such feedback into robust training objectives for LLMs that align with users' true preferences.
More importantly, balancing novelty and relevance -- two usually competing objectives -- is crucial for exploration in recommendation systems as relevant novel content drives sustained user satisfaction. 
%This balance has proven difficult to achieve \cite{wang2023fres}.
Initial experiments with the hierarchical planning \cite{wang2024llms} framework, using an LLM as a novelty 
model to identify novel interest clusters and subsequently retrieve relevant items, demonstrated the potential of this approach. However, aligning the novelty model's predictions with user preferences remains challenging.  Directly fine-tuning with more users' interaction history data yielded neutral results and raised concerns about memorization and loss of novelty. Attempts at RLHF \cite{ouyang2022training} with a reward model also proved unsuccessful as it undermined the controlled generation capability (see in Sec. \ref{sec:preliminary}). 

%\yl{this paragraph and preliminary are similar, need to merge, leaning merging into preliminaries}
%\yl{how about just say wang et al had initial success in achieving high novelty with good relevancy, but much headroom remains in improving user satisfaction --> hence the new approach in the next paragraph}

To address these challenges, we propose a novel, decomposed approach that leverages two specialized LLMs for high-level planning: a novelty model and an alignment model. 
%\jw{Recognizing the limitations of noisy, implicit individual user data, we shift our focus to collective user behaviors, aggregating preference signals at the LLM-generated cluster level. By analyzing collective clicks, dwell times, and other interactions with the recommended items, we can discern broader patterns of user interest and satisfaction, mitigating the impact of individual noise.} 
To balance novelty and relevance, the alignment LLM is trained specifically to evaluate and rate the predictions of the novelty model based on observed user feedback. This separation allows for the independent optimization of novelty generation and preference alignment. Moreover, to further improve the system's ability to generate relevant novel predictions, we scale inference-time compute by generating multiple independent predictions from the novelty model using a high temperature setting. The alignment model then acts as a selector, choosing the most user-aligned outputs from the novelty model. This combination of specialized models, training signals derived from collective user behaviors, and repeated sampling significantly increases the likelihood of generating recommendations that are both novel and relevant.

In summary, this paper presents a system that has been \textbf{deployed} on a commercial short-form video recommendation platform serving billions of users. The key contributions are: (1) \textbf{Collective User Feedback Alignment}: We introduce an LLM-based alignment model specifically trained to evaluate the novelty model's predictions based on collective user behaviors. By aggregating implicit signals(e.g. clicks and dwell time) for interest clusters transition across many users, we enable the system to learn user preferences with reduced noise and bias. (2) \textbf{Inference-Time Scaling}: We demonstrate the effectiveness of repeated sampling at inference time, allowing the alignment model to select the most relevant predictions from a diverse set of candidates generated by the novelty model, thereby improving exploration. (3) \textbf{Decomposed Novelty and Preference Modeling}: We propose a novel paradigm that decouples novelty generation and preference modeling into two specialized LLMs. This separation enables independent optimization for each objective. Consequently, it directly addresses the core challenge of balancing novelty with relevance via specialized models, leading to a significantly improved operating curve for user interest exploration. 

\section{Related Work}
This research builds upon two primary streams of existing work: the application of LLMs to recommendation systems and the ongoing efforts to improve recommendation exploration.

\smallskip
\noindent\textbf{LLMs for Recommendation Systems}. 
The advances in LLM capabilities have recently drawn a lot of attention to their potential in recommendation systems \cite{bao2023tallrec,geng2023vip5,hou2023large,li2023gpt4rec,liu2023chatgpt,wang2024fresh}. One promising direction involves augmenting traditional recommendation models with LLM-powered feature engineering, including supplementary textual features or embeddings that encode world knowledge \cite{xi2024towards,ren2024representation}. Another approach focuses on directly generating recommendations using LLMs; e.g., \citeauthor{hou2023large} and \citeauthor{gao2023chat} have experimented with prompting off-the-shelf LLMs to produce ranked lists of recommendations. Meanwhile, there are also work involving fine-tuning LLMs \cite{singh2024better,bao2023tallrec, lin2024rella} to better align them with the recommendation domain, whether through incorporating domain-specific knowledge, generating new tokens, or predicting user preferences for specific user-item pairs. However, few of these methods are truly equipped to handle query-per-second (QPS) requirements of real-time applications. \cite{wang2024large} addresses this by employing LLMs as data augmentation tools for conventional recommendation systems during training, thereby boosting performance without incurring additional serving costs. 

\smallskip
\noindent\textbf{Recommendation Exploration}. Improving user interest exploration is key to broadening preferences and fostering long-term engagement \cite{chen2021values,chen2021exploration, su2024long}. However, a key challenge lies in the inherent closed-loop nature of existing recommendation systems~\cite{chaney2018algorithmic,mansoury2020feedback,wang2023fresh}. Training data is primarily derived from past user-item interactions, limiting the system's ability to explore truly novel interests. While methods like PIE \cite{mahajan2023pie} offer improvements through user-creator affinity and online bandit formulations, they remain confined by the system's internal knowledge \cite{chen2021values}. Building on the LLM-powered hierarchical planning architecture \cite{wang2024llms}, which guides user interest exploration at the cluster level, we focus on enhancing its performance through user feedback alignment. Our work investigates the integration of effective user feedback signals into LLMs for recommendation systems.

% \section{Method}
\section{Preliminaries}
\label{sec:preliminary}

\smallskip
\noindent\textbf{Hierarchical Planning Paradigm.} In the hybrid hierarchical planning paradigm \cite{wang2024llms}, LLMs focus on high-level planning by predicting novel user interests at the interest cluster level. Interest clusters are topically coherent item clusters generated from item metadata and content embedding \cite{chang2024cluster}. To provide the LLM with domain knowledge of our system, we fine-tuned the LLM using the novel interest transition patterns mined from users' interaction history. 

%Figure \ref{fig:prompt} is a LLM training example used during finetuning. \yl{decide what to do with this reference}

\begin{figure}[t]
% \vspace{-0.1in}
\centering
\includegraphics[width=0.5\textwidth]{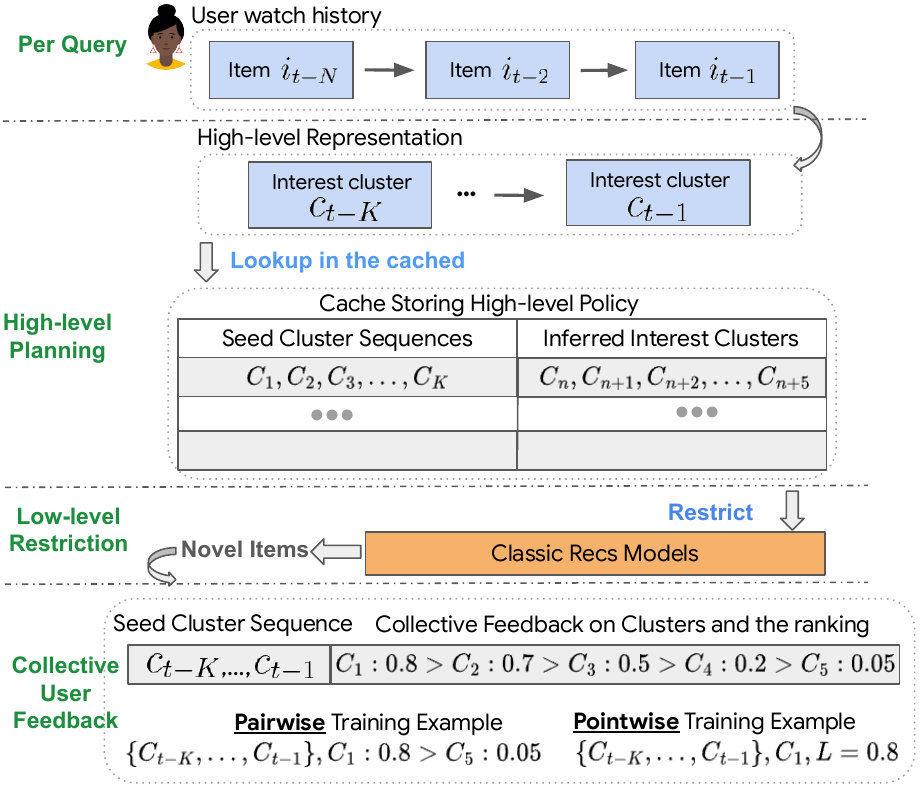}
%\vspace{-0.1in}
\caption{Hierarchical planning paradigm: the novelty LLM performs high-level planning for novel interest transitions, which are used to restrict the predictions of classic recommender models, and user feedback on these novel recommendations is aggregated to train a separate alignment LLM.
}
%\vspace{-0.1in}
\label{fig:hierarchical_planning}
\end{figure}

As illustrated in Figure \ref{fig:hierarchical_planning}, during the high-level planning, given a user's recent interaction history, represented as a sequence of $K$ clusters $S_u$ (i.e., $|S_u| = k$), the LLM predicts the next novel cluster $C_n$ for this user. 
Because online serving the LLM for a billion-user system is prohibitively costly, we pre-compute and store potential next interest transitions for all combinations of sampled k clusters $\mathcal{\textbf{S}} = \{ S \mid S \subseteq \{C_1, C_2, \dots, C_N\}, |S| = k \}$. During online serving, a user's history is mapped to the corresponding pre-computed novel interest through looking up the precomputed interest transitions. At the lower level, a conventional, transformer-based sequential recommender backbone  handles the computationally intensive task of item-level selection. However, instead of searching the entire item space, the backbone is constrained to recommend items only within the novel interest clusters $C_n$ identified by the LLM. This constraint combines the personalization capabilities of the backbone with the novelty-seeking behavior of the LLM, leading to a personalized recommendation experience enriched with serendipitous discoveries.

We've launched this user interest exploration paradigm to the production recommendation system, which resulted in a rare combination of high novel item ratio and user satisfaction gain. The lightweight finetuning (<8k training examples) was key to preserving the LLM's pre-trained knowledge while imparting an understanding of our users' interaction patterns. 

\smallskip
\noindent\textbf{Limitation.} The lightweight finetuning has limitations: 1) The 8k training examples represented a limited view of the behavior of our large user base.  2) For the cluster combinations that are hard to reason, LLM has low prediction confidence, indicated by the novel interest predictions that don't have a logical connection to the users' existing interests. This hurts the relevancy of the recommendation and, consequently, user satisfaction. 

To improve the relevancy of the novel interest prediction, we initially tried to increase the number of training examples. However, novelty metrics in A/B testing didn't show sensitivity to this change. Furthermore, due to the LLM's tendency to repeat training data (our analysis showed a 40\% chance of repetition during inference), scaling to more training examples mined from the user history risked reinforcing the system feedback loop, impairing LLM's ability to make novel recommendations.

To align the LLM with user preference without amplifying the system feedback loop, we leverage live-traffic users' feedback to LLM's own recommendations, such as clicks, dwell time and repeated interaction, which is independent of the system behavior. We first tried the classic RLHF setup: RL fine-tune the novelty LLM directly with a reward model trained with user preference. However, this always resulted in the model quickly collapsing: 1) loss of controlled generation: after 5k steps, the LLM's chance of predicting in the correct format drops from 99+\% to 2\%; 2) Reward hacking: the model learned the high reward words, e.g., 'cat', 'BTS', 'toys', etc, and frequently predicts those words. While RLHF is effective for free form text generation in conversation settings, it proved insufficient in structured tasks with strict format and content vocab requirements -- the reward model cannot capture the nuanced task requirements and guide the RL finetuning process accordingly.

\section{Method}

To address the challenges in classic RLHF, we introduce an inference-time scaling method \cite{brown2024large} with a decoupled dual-specialization modeling approach. Instead of directly fine-tuning the policy model (i.e., the novelty model for planning the next cluster) through SFT or RLHF, we first performs independent sampling from the novelty model. This generates a diverse set of candidate interest clusters. Subsequently, the best-of-n clusters are selected using a separate alignment model trained on collective user feedback based on their likelihood to resonate with users.

% Uncovering novel yet relevant (i.e., engaging) item clusters with the high-level policy model is critical for overall recommendation performance. Previous work \cite{wang2024llms} explored fine-tuning a single LLM for controlled generation and basic user behavior alignment. However, scaling training data to achieve stronger alignment with user preferences can lead to overfitting and the inadvertent reinforcement of existing biases when predicting novel interest clusters.

This section details our design, demonstrating: (1) the methodology for collecting and transforming implicit user feedback from interactions with the recommendation system into fine-tuning signals for the alignment model; and (2) a top-n selection strategy and inference scaling approach that simultaneously optimizes for both relevance and novelty with minimum latency impact, showcasing its practical applicability in large-scale real-world recommendation systems.

\subsection{Preference Alignment on User Feedback}

\smallskip
\noindent\textbf{Aggregating Collective Human Feedback.}
Through per-query logging inside our LLM-powered recommender serving live traffic (detailed in Section \ref{sec:preliminary}, `Novelty model' hereafter), we collect users' preferences on LLM's predictions. Specifically, for each predicted cluster $C_n$, we log the cluster sequence  \{$C_1$, ..., $C_{K}$\} used to represent the user, and the user's feedback on $C_n$ (e.g., positive playback, like, share, skip, etc). We then aggregate the feedback for each (\{$C_1$, ..., $C_{K}$\}, $C_n$) pair, resulting in user preference training examples denoted as (\{$C_1$, ..., $C_{K}$\}, $C_n$ , $L_{(1,k),n}$). Here, $L_{(1,...,k),n}$ represents the aggregated user feedback score (e.g. like rate, share rate) for this particular interest cluster transition – that is, serving interest cluster $C_n$ to a user with historical viewing pattern represented by \{$C_1$, ..., $C_{K}$\}). 

We then post-process the aggregated feedback to: 1) normalize the feedback score, which can be skewed towards very small values because the feedback signals, e.g. like, share, etc, are sparse. 
%We've tried different normalization strategies and the details can be found in the Appendix. 
2) filter cluster transition pairs with little user feedback. 3) round the feedback score to a fixed interval to account for margin of error in the aggregated stats.

Besides the aforementioned \textit{pointwise training example} (\{$C_1$, ..., $C_{K}$\}, $C_n$ , $L_{(1,k),n}$), we also tested pairwise training examples: we rank the different $C_n$ for a cluster sequence  \{$C_1$, ..., $C_{K}$\} by the aggregated feedback score, and we create training examples by sampling contrastive $C_n$ pairs as labels. Pairwise training examples require neither normalization nor picking a threshold for positive labels. We can also generate more training (K-choose-2 vs K) examples per cluster sequence.

\smallskip
\noindent\textbf{Alignment Reward Model Training.} To align with collective user feedback, we trained an "alignment model"(a reward model) to score the users' affinity to $C_n$ given their watch history. The alignment model is training using a cross-entropy loss between its prediction and the user’s actual aggregated engagement metric (i.e., positive playback rate). This alignment model is an LLM with the last layer being a linear projection layer.  

% \jw{are we using the same loss function for pairwise and pointwise?}
% \yl{Jianling, let's update this prompt using the correct label and model prediction score}
\begin{figure}[t]
%\vspace{-0.15in}
\centering
\includegraphics[width=0.5\textwidth]{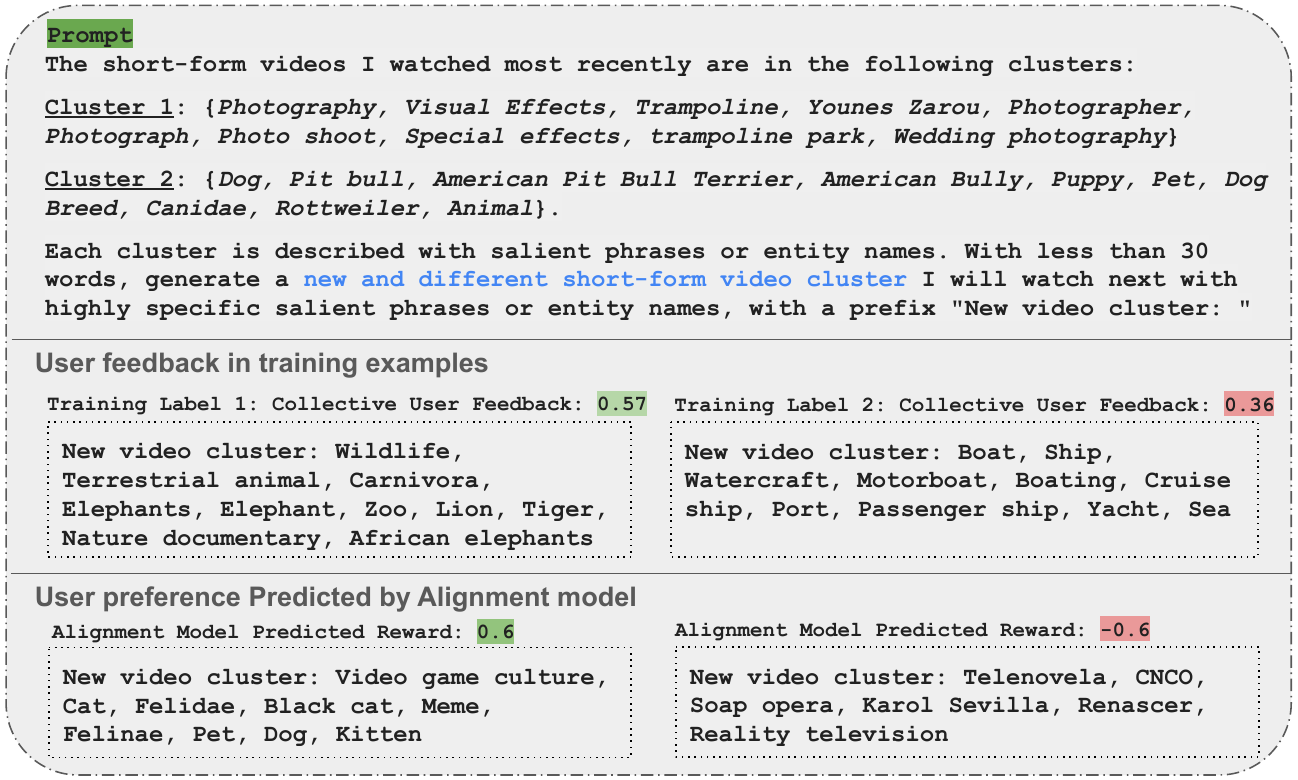}
%\vspace{-0.1in}
\caption{The alignment model trained with collective user feedback can effectively predicts user preference over new labels.}
%\vspace{-0.1in}
\label{fig:prompt_compare}
\end{figure}

In Figure \ref{fig:prompt_compare}, we showcase a sample prompt describing users with \{photography, Visual Effects, Special effects\} and \{dogs\} interest clusters(assuming $K=2$). Collectively, those users prefer label 1(\{wildlife, nature documentary\}) over label 2  (\{boats\}) as expressed in the feedback scores. Given two new labels, the trained alignment model also effectively assigns high preference score to \{cats, video game, internet meme\} over less relevant next interest cluster. These intuitive examples demonstrate the feasibility and potential of the alignment training.

\subsection{Inference Scaling with Best-of-N User Alignment}
% In the previous effort \cite{wang2024llms}, every candidate from the high-level policy model is served to downstream systems. 
We use the user alignment model as surrogate for user preference to critique the relevancy of the novel clusters predicted by the novelty LLM, which itself is lightly fine-tuned with users' interaction histories. To increase the chance of predicting a novel cluster that is more aligned with user preference, we repeatedly and independently sample 5 times more  predictions from the novelty LLM with high temperature, and then rank the predictions using the alignment model and pick the top k where k is the number of clusters served by the production system. Because the novelty LLM sampling, reward model scoring, and the best-of-n selection all happens offline, and we serve the same number of clusters in live traffic, there is no latency impact, and the additional cost of scaling up inference is amortized across offline bulk inference runs.

Maintaining the novelty of predictions is crucial for effective user interest exploration. The repeated sampling of the novelty LLM improves the reasoning quality and maintains the prediction novelty while the alignment model selects the predictions users may prefer. This dual LLM setup avoids the challenge of teaching an LLM both novelty and relevancy -- two competing objectives that can risk catastrophic forgetting. By evaluating the novelty prediction using an LLM aligned with user feedback, we improve the exploration efficiency by demoting the predictions that may result in lower user satisfaction.

\section{Live Experiments}
\label{sec:live_exp}
%In this section, we examine how effectively a hierarchical planning with LLM design improves user interest exploration. 
\subsection{Experimental Setup}
Our live experiments were conducted on a commercial short-form video recommendation platform serving billions of users. While we employed Gemini \cite{team2024gemini} for both the novelty and alignment models, the fine-tuning process and pipeline are designed to be adapted to other LLMs. The high-level planning recommends novel interest clusters based on a user's historical interest cluster sequence of length $K=2$, and the system is designed to accommodate larger K values in the future through a sparse table implementation.

\begin{figure}[t]
\vspace{0.1in}
\centering
\includegraphics[width=0.48\textwidth]{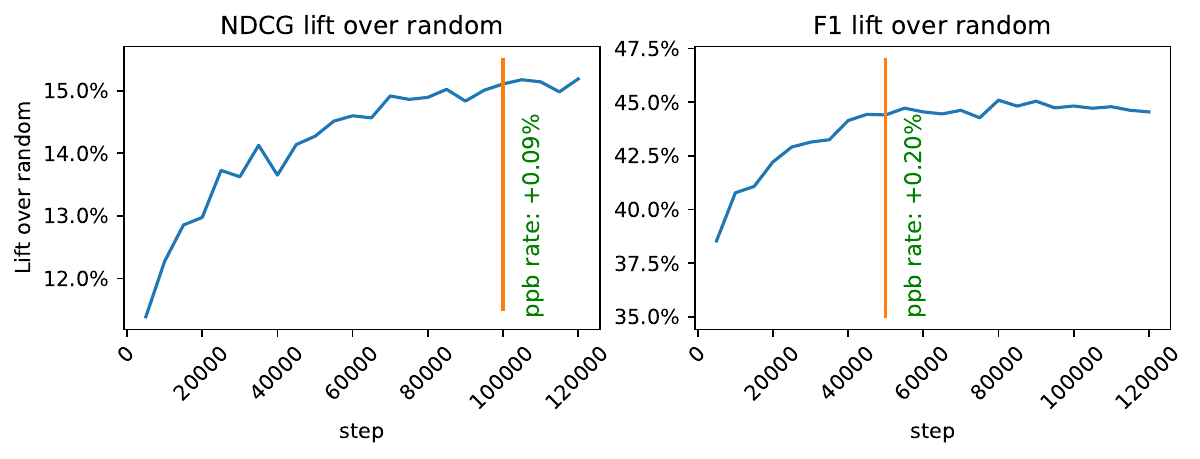}
%\vspace{-0.1in}
\caption{Alignment Model Finetuning and Evaluation.}
%\vspace{-0.1in}
\label{fig:offline_result}
\end{figure}

\begin{figure*}[ht!] % Added [ht!] for better placement
\vspace{-0.8cm}
    \centering % Added \centering for horizontal centering of the figures
    \begin{subfigure}[b]{0.3\textwidth} 
        \includegraphics[width=\linewidth]{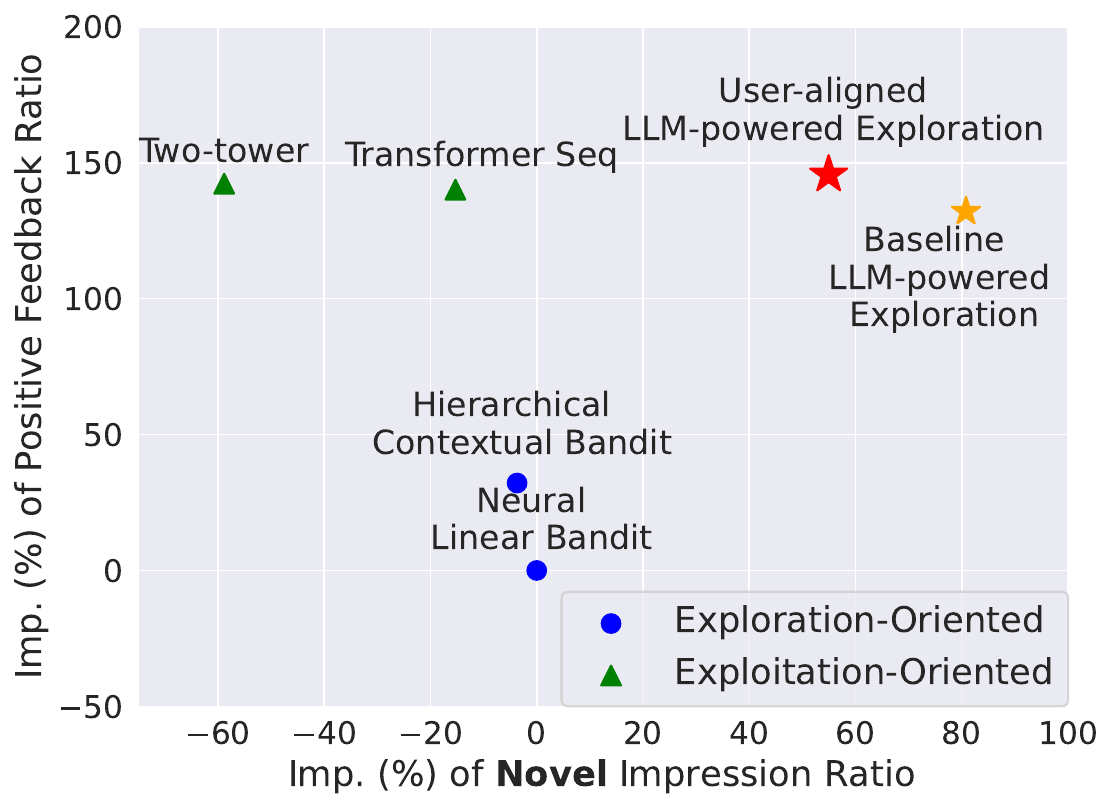}
        \caption{Novelty VS Quality}
        \label{fig:sub1} % Added label for referencing
    \end{subfigure}%  <-- Important: No space here
    \hfill % Fill the space with an empty element
    \begin{subfigure}[b]{0.23\textwidth} 
        \includegraphics[width=\linewidth]{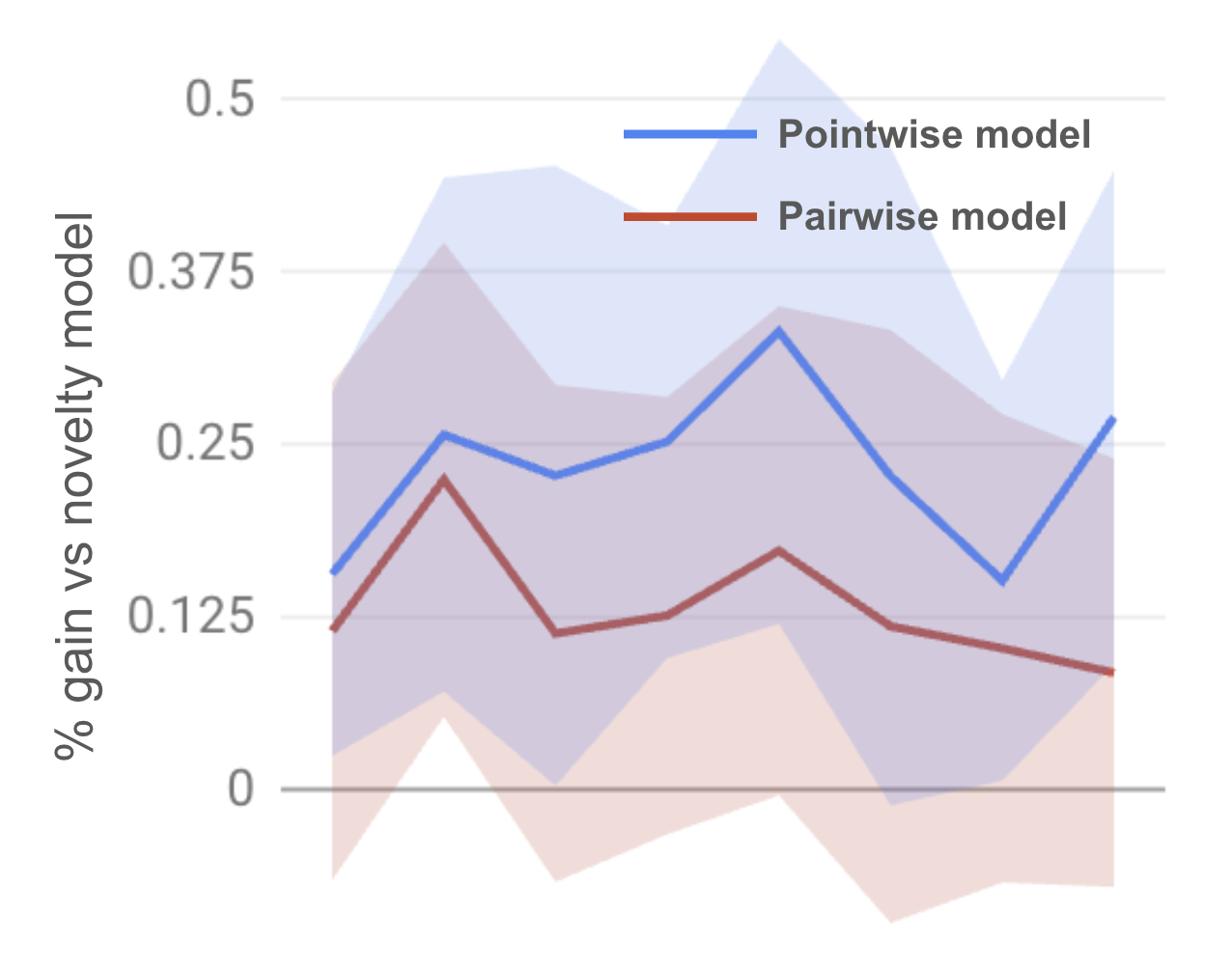}
        \caption{Unique User-Cluster}
        \label{fig:sub2} % Added label for referencing
    \end{subfigure}%  <-- Important: No space here
    \hfill % Fill the space with an empty element
    \begin{subfigure}[b]{0.23\textwidth} 
        \includegraphics[width=\linewidth]{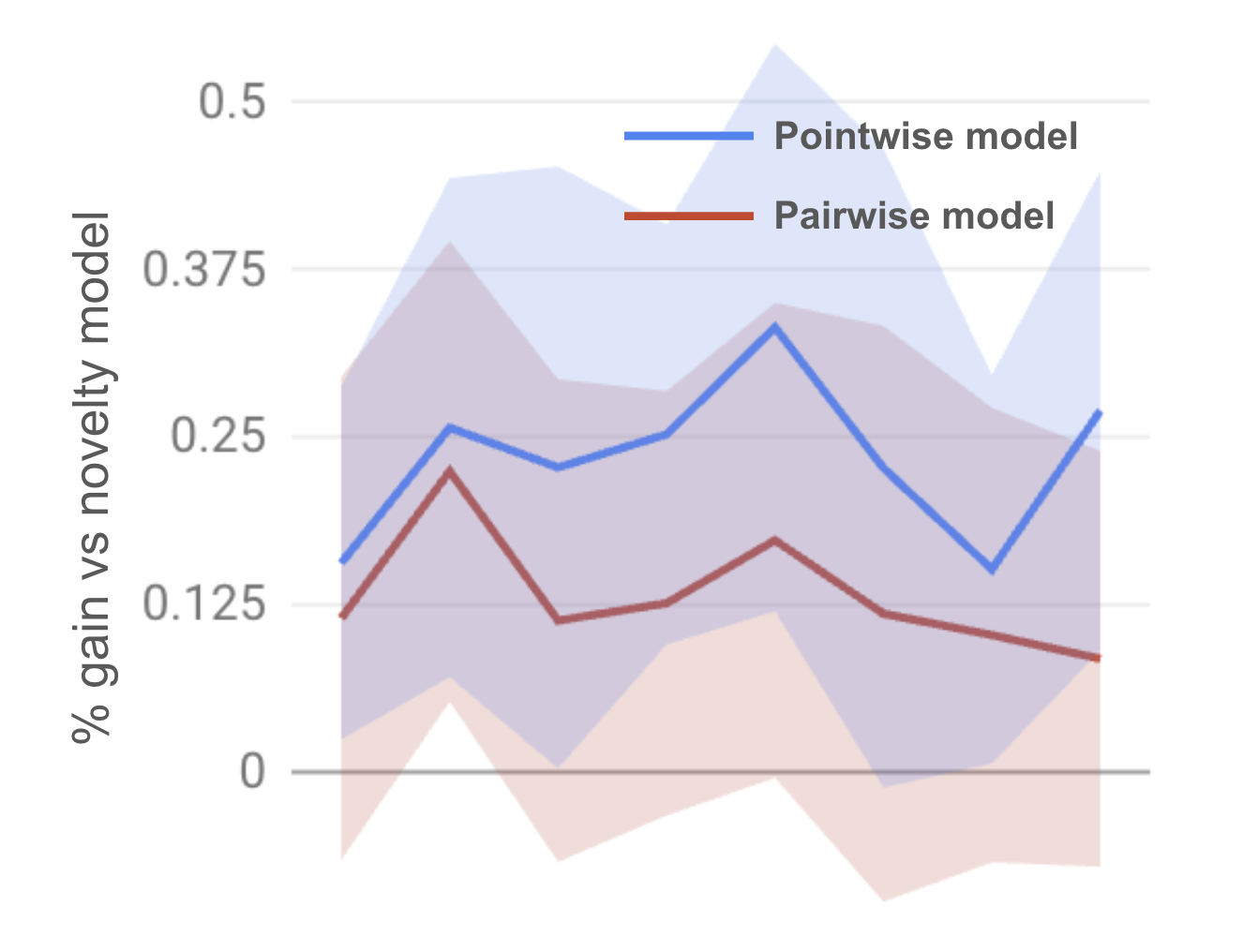}
        \caption{Pos Playback Rate}
        \label{fig:sub3}
    \end{subfigure}%  <-- Important: No space here
    \hfill % Fill the space with an empty element
    \begin{subfigure}[b]{0.23\textwidth} 
        \includegraphics[width=\linewidth]{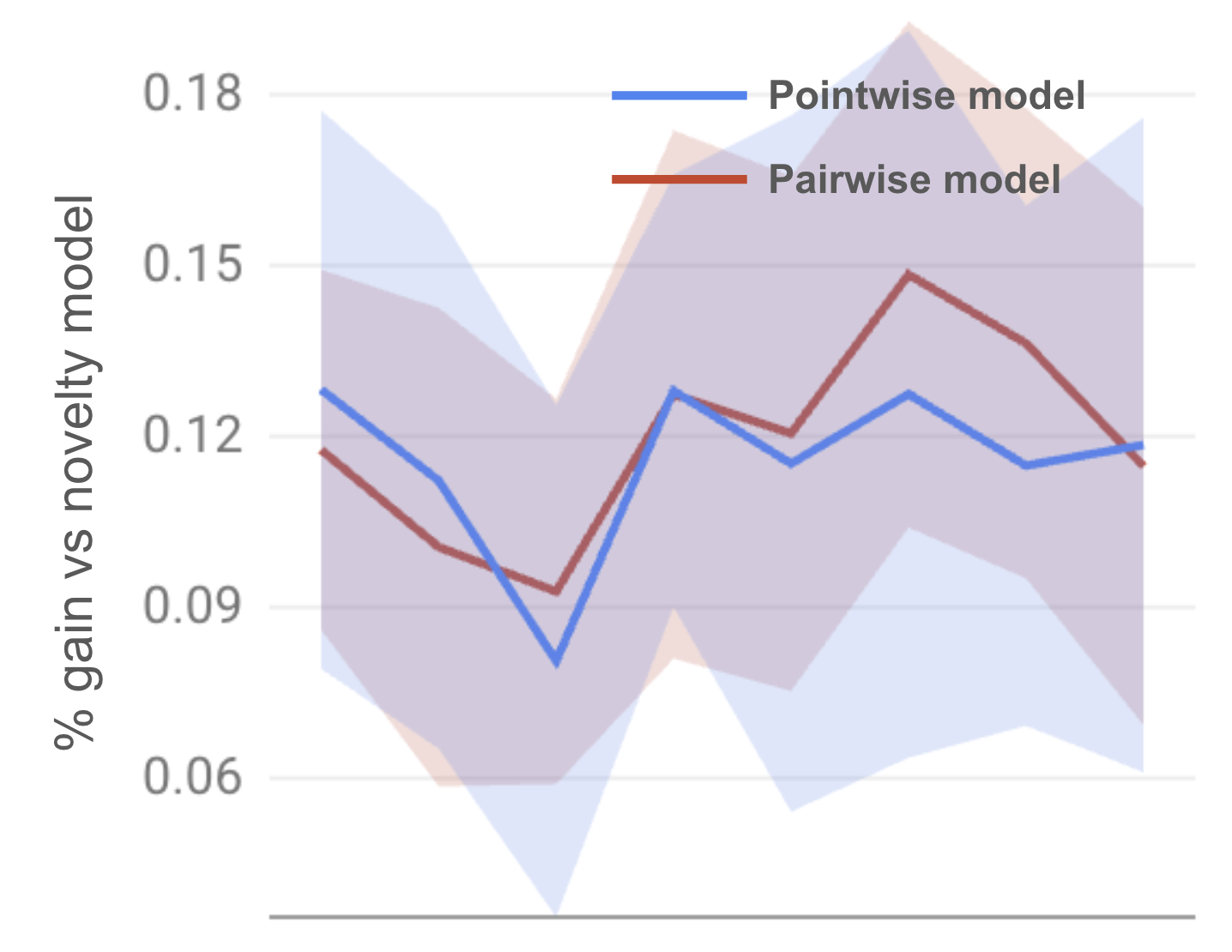}
        \caption{Completion Rate}
        \label{fig:sub4} % Added label for referencing
    \end{subfigure}%  <-- Important: No space here
    \caption{(a) The proposed method still recommends the highest percentage of novelty compared to the rest of the system. (b)(c)(d) Compared to the novelty model baseline, the alignment model further expands users' interest with higher user satisfaction.}
    \label{fig:supi}
    \label{fig:live}
    %\vspace{-0.5cm}
\end{figure*}

\smallskip
\noindent\textbf{Baseline.} 
Besides comparing to the baseline novelty model without user alignment\cite{wang2024llms}, we also compare the proposed method to existing production models: (1) \textbf{Exploration-oriented} models include: \textit{Hierarchical contextual bandit} \cite{song2022show} obtain the next clusters through a tree-based LinUCB; \textit{Neural linear bandit}-based DNN model \cite{su2024long} to predict the next novel cluster. 
Although these models are tailored to explore user interests, they are trained on interest transitions existing in the system and therefore are still subject to the feedback loop. 
(2) \textbf{Exploitation-oriented} models include a regular \textit{two-tower} model \cite{yang2020mixed} and \textit{transformer-based} \cite{chen2019top,shaw2018self} sequential model trained on all positive user feedback. Our live experimental results demonstrate our proposed method can lead to recommendations that are more novel and of better quality compared to these existing models. 

\subsection{Model Finetuning and Offline Evaluation}
We used offline metrics to guide the alignment model training, checkpoint selection, and hyper-parameter searching (e.g. score normalization strategy). Offline evaluation is done on a holdout set of interest cluster sequences, the novel interest transitions and user's feedback scores. During evaluation, the alignment reward model scores and ranks the interest cluster transitions for each input cluster sequence. We compare this model-generated ranking against the ground-truth ranking from the user feedback. Performance is measured using F1@K (i.e., the harmonic mean of precision and recall), and NDCG@K metrics, with K being the number of interest clusters served in live traffic.

As shown in Figure \ref{fig:offline_result}, the offline metrics improve consistently over a random baseline  throughout the alignment model's training process. These results underscore the importance of incorporating user feedback alignment into our inference scaling approach. Furthermore, the offline evaluation guided the hyper-parameter tuning, allowing us to optimize the reward model's performance and prevent overfitting. In live A/B experiments, we deployed two arms: one favorable arm using an alignment model trained for 50,000 steps (where F1 converged in offline evaluation as shown in Figure \ref{fig:offline_result}), and another arm using an alignment model trained for 100,000 steps beyond the favorable converging point as comparison. We observed significantly improved user satisfaction with the favorable arm, as evidenced by a larger positive playback rate gain – indicating better alignment with user preferences. This finding is consistent with our hypothesis that extensive training beyond the convergence point can lead to overfitting. While NDCG encourages the model to reproduce the exact ranking from user feedback, F1@K focuses on the model's ability to identify the top-K most relevant clusters, which is more crucial for our top-n selection task. Memorizing the exact rankings is unnecessary and potentially detrimental to the exploration of novel and engaging recommendations.

%We compare with the following baselines which are already deployed in production: Type(1) \textbf{Novelty-enhanced sequence recommender} ~\cite{chen2021values}. Such model is trained with labels from both positive and novel items whose clusters have not appeared in user's consumption history before. (2) \textbf{Hierarchical contextual bandit} \cite{song2022show} based on the hierarchical clusters introduced in \ref{sec:preliminary}. Such model explores user's interests through a tree-based LinUCB to obtain the next clusters, from which the sequential model is then used to restrict the retrieval to items. (3) \textbf{Neural linear bandit}-based DNN model \cite{su2024long} to predict the next novel cluster, from which the same sequential model is then used for item-level policy, similarly as hierarchical contextual bandit.(4) Two-tower model and transformer-based sequential model trained for exploitation. Though this models are tailored for extend user interest, they are trained on interest transitions existing in the system and therefore still fall into the feedback loop. (5)  And our live experimental results demonstrate our proposed method can lead to recommendation which are more novel and in better quality compared to the existing models. 

\subsection{Results and Analysis}
This section shows our method simultaneously improves recommendation novelty and user satisfaction, outperforming baselines in engagement and exploration, and details the benefits of its production-deployed pointwise labeling strategy.

\smallskip
\smallskip
\noindent\textbf{Novelty and Quality.} In Figure \ref{fig:supi} (a), we compare the proposed method with various baseline models currently in production. Using the performance of Hierarchical contextual bandit \cite{song2022show} as the base, we measure improvement of novelty and quality of other models in our system. Specifically, we plot the increase in the novel impression ratio (impressions from interest clusters the user has never interacted with) to highlight recommendation novelty (x-axis), and the increase in positive playback rate to demonstrate recommendation quality (y-axis). We observed that aligning the novelty model with user preference results in higher users' positive playback ratio at a slight cost of novelty. Nonetheless, the proposed method still has the highest novel impression ratio compared to the rest of the system. Additionally, our method achieves significantly better quality than existing exploration-oriented methods, even surpassing the exploitation-oriented methods. It is rare in recommendation systems to achieve high novelty and user satisfaction simultaneously. This means through user feedback alignment, we moved our model to a more optimal point in the operation curve -- over user satisfaction and engagement improved while the novelty is still the highest in the system.

\smallskip
\noindent\textbf{Increased User Satisfaction.} In Figure \ref{fig:live}(c), (d), the x-axis represents the experiment periods (the exact dates are redacted), and the y-axis shows the relative percentage difference between the experiment and control.  We observed an increase in the positive playback rate and the completion rate of the recommended content, indicating an increased user satisfaction on the platform. %  illustrates the user metrics from our live experiments. 
%The collective experiment metrics gains shows that our method successfully broadened user interest with diverse and novel recommendation, leading to increased user satisfaction. 
%This underscores the quality and relevance of the recommended novel content. 

\smallskip
\noindent\textbf{User Interest Exploration.} To measure if the recommender encourage users to explore new interests, we use unique engaged user-cluster (\textbf{UEUC}), which tracks the number of unique user-cluster engagement pairs. Figure \ref{fig:supi}(b) shows that our proposed user feedback alignment method not only improves the user satisfaction but also improves the number of user interests. This means our method improves the exploration efficiency. We also observed UEUC is higher for more active users, potentially because the reward model aligns more closely with the preferences of core users who contribute a larger portion of the user feedback training data.

\smallskip
\noindent\textbf{Pairwise vs Pointwise Label.} The live experiment results shown in Figures \ref{fig:live}(b), (c), and (d) demonstrate a performance comparison between alignment models trained with pairwise labels and those trained with pointwise labels. Both models positively impact user's interest size and satisfaction, with the pointwise model slightly outperforming the pairwise model. This indicates normalizing users' feedback per the feedback's prior helps. Pairwise model learns the relative rank of the novel clusters and its scoring of new cluster may be uncalibrated, thus negatively impacting the performance. We also observed that the pointwise model training is 2x faster. Hence the pointwise model was deployed to production.

\section{Conclusion}
In this paper, we advanced the hierarchical planning paradigm for LLM-powered large-scale recommendation systems by decoupling high-level planning into two specialized models: one focused on generating novel interest candidates and another focused on aligning these candidates with user feedback. We share our successful approach to improving alignment using collective user feedback gathered from LLM-powered recommendation systems.  Live experiments on a large-scale recommendation platform demonstrate that our proposed method enhances exploration efficiency while simultaneously increasing user engagement.

\bibliography{custom}

\end{document}